\documentclass[preprint,12pt,authoryear]{elsarticle}

\usepackage{amssymb}

\usepackage{graphicx}

\journal{Icarus}

\begin{document}

\begin{frontmatter}



\title{Assessing the contribution of Centaur impacts to ice
giant luminosities}


\author{Sarah E. Dodson-Robinson}

\address{University of Delaware, Physics and Astronomy
Department, 217 Sharp Lab, Newark, DE 19716, USA}

\ead{sdr@udel.edu}

\begin{abstract}
Voyager 2 observations revealed that Neptune's internal luminosity is an
order of magnitude higher than that of Uranus. If the two planets have
similar interior structures and cooling histories, Neptune's luminosity
can only be explained by invoking some energy source beyond
gravitational contraction. This paper investigates whether Centaur
impacts could provide the energy necessary to produce Neptune's
luminosity. The major findings are (1) that impacts on both Uranus and
Neptune are too infrequent to provide luminosities of order Neptune's
observed value, even for optimistic impact-rate estimates, and (2) that
Uranus and Neptune rarely have significantly different impact-generated
luminosities at any given time. Uranus and Neptune most likely have
structural differences that force them to cool and contract at different
rates.
\end{abstract}


\end{frontmatter}


\section{Introduction}
\label{sec:intro}

While the ice giants may have similar interior structures
\citep[e.g.][]{podolak95, fortney10}, their internal luminosities differ
by a factor of 10. From Voyager 2 IRIS radiometer observations,
\citet{pearl91} calculated an internal luminosity of $\log L/L_{\odot} =
-11.024$ for Neptune, while \citet{pearl90} found an internal luminosity
of $\log L/L_{\odot} = -12.054$ for Uranus. The $2.7 M_{\oplus}$ mass
difference between the two planets is not enough to explain the
luminosity difference: the internal power generated per unit mass is
$3.22 \times 10^{-7}$~erg~g$^{-1}$~s$^{-1}$ for Neptune and $3.92 \times
10^{-8}$~erg~g$^{-1}$~s$^{-1}$ for Uranus \citep{pearl90, pearl91}.
Multiple theories explaining the energy balance of the ice giants have
been put forward, including stable stratification in Uranus' interior
\citep{podolak90}, early and efficient heat transport by baroclinic
instability in Uranus \citep{holme94}, and efficient capture of strongly
interacting dark matter by Neptune \citep{mitra04, adler09}.

One energy source that has not been investigated in connection with ice
giant energy balance is impact heating. Given a sufficient supply of
Centaurs\footnote{While we use the word ``Centaur'' loosely to describe
any object that may impact Uranus or Neptune, \citet{jewitt09} defines
Centaurs as comets whose dynamics are controlled by perihelion and/or
aphelion interactions with giant planets, such that perihelia $q$ and
semimajor axes $a$ are in the range $5.2 < (q,a) < 30.0$~AU. The Minor
Planet Center website,
minorplanetcenter.net/blog/asteroid-classification-i-dynamics/, defines
a Centaur as an asteroid with $q > 5.2$~AU and $a < 30.0$~AU.}, impacts
onto Neptune could be frequent enough to boost Neptune's luminosity to
observed values. Indeed, meteoroid impacts onto the moon generate
flashes of optical light, first observed by \citet{dunham99}. Energy
deposited in ice giant atmospheres by Centaurs that penetrate the
photosphere would not be released instantly to space, as in the case of
lunar meteoroid flashes, but would instead be radiated away on a $\sim
100$-year timescale \citep{conrath90}. This paper explores the
possibility that Centaur impacts may contribute significantly to ice
giant luminosities.

The investigation begins with an order-of-magnitude calculation of the
typical Centaur size required to produce Neptune's luminosity with
impacts alone, treating impacts as a steady-state process. Next, we
explore different impact rates and break the steady-state assumption,
treating impacts as a stochastic process. Section \ref{sec:sizedist}
contains estimates of the total number of Centaurs, which we use as a
scaling factor for published impact rates.
Section \ref{sec:luminosity} describes a Monte Carlo approach to
computing a cumulative probability distribution of planet luminosity.
Results and conclusions are presented in Section \ref{sec:results}.

\section{Impact-Induced Luminosity: Order-of-Magnitude Estimate}
\label{sec:oom}

To get a basic idea of how much impacts contribute to ice giant
luminosities, we assign a constant value $\dot{M}$ to each planet's
accretion rate and assume a constant accretion-generated luminosity. The
impact-generated luminosity is then
\begin{equation}
L_{\rm imp} = \frac{G M \dot{M}}{R},
\label{eq:accretion}
\end{equation}
where $M$ is the planet mass and $R$ is the planet radius.  Equating
$L_{\rm imp}$ with Neptune's present luminosity requires $\dot{M}
= 4 \times 10^{17}$~g~yr$^{-1}$.
Based on simulations of diffusion from the Kuiper Belt to the inner
Solar System, \citet{levison97} found that comets impact Uranus and
Neptune slightly more than once per thousand years. To deliver the
average $\dot{M}$ quoted above, most of the impacting Centaurs
with a density of $\sim 1$~g~cm$^{-3}$ would have to have radii over
40~km. Such a large average Centaur size can be ruled out by crater
observations; for example, \citet{stern00} calculated that the largest
craters detected on Triton were created by impactors with radii of
1-5.5~km. The occultation surveys of \citet{roques06} and
\citet{schlichting09, schlichting12} also indicate a Centaur/Kuiper Belt
Object size distribution heavily biased toward sub-kilometer bodies.

In the steady-state scenario where $L_{\rm imp}$ is constant and the
\citet{levison97} impact rate applies, impacts clearly cannot drive
Neptune's internal heating. Explaining Neptune's luminosity with impacts
alone requires one of two scenarios: (1) a substantially higher impact
rate, which is possible if \citet{levison97} underestimated the total
number of Centaurs; or (2) a recent giant impact that has driven
Neptune's luminosity to an above-equilibrium value. The rest of this
paper examines scenarios (1) and (2).

\section{Total number and size distribution of Centaurs}
\label{sec:sizedist}

Determining the frequency and energy of impacts on ice giants requires
knowing both the total number of Centaurs and their size distribution.
The number of Centaur detections is too small to reconstruct a size
distribution based on observations alone: only 7 Centaurs met the
``secure orbit'' standards used by the Deep Ecliptic Survey team to
compute a debiased $H$-magnitude distribution \citep{adams14}.
Fortunately, Centaurs have short dynamical lifetimes, so their size
distribution is a relic of their source population. The cold Kuiper Belt
\citep[e.g.][]{holman93, levison97, fraser10, volk11}, the Neptune
Trojans \citep{horner10}, the inner Oort cloud \citep{emelyanenko05,
kaib09, brasser12, volk13, delafuentemarcos14, fouchard14}, the Plutinos
\citep{morbidelli97, disisto10}, and the scattered disk
\citep{disisto07, volk08, volk13} could all be Centaur sources. However,
no empirical information exists on the size distribution of objects in
the Oort cloud, and \citet{fraser10} find that the scattered disk is not
populous enough to explain the observed influx of comets into the inner
Solar System. \citet{doressoundiram05} also show that Centaur colors are
not consistent with an origin in the scattered disk. Moreover,
\citet{schlichting13} show that the cold Kuiper Belt and scattered disk
objects have size distributions that follow the same functional form,
only with different maximum sizes.  Calculations presented here are
based on the cold Kuiper Belt size spectrum of \citet{schlichting13},
which is a close match to the size spectrum of Saturnian satellite
impactors inferred from the cratering record \citep{minton12}.
\citet{schlichting13} used a combination of theoretical coagulation
models, occultation surveys, and observations of large KBOs to constrain
the size spectrum.



\begin{figure}
\centering
\includegraphics[width=0.8\textwidth]{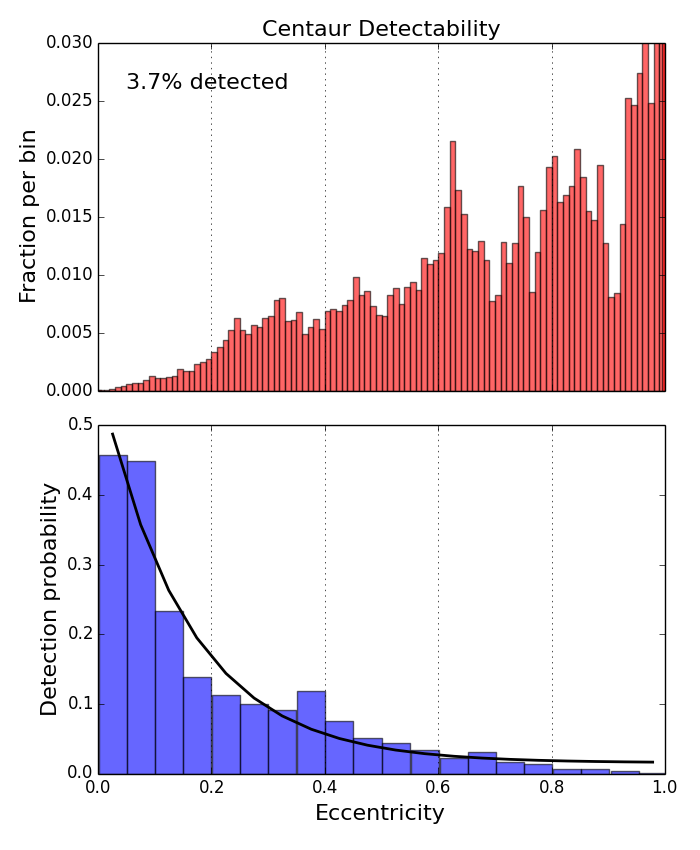}
\caption{{\bf Top:} Time-average of Centaurs per eccentricity bin from
the 100~Myr Centaur dynamics simulation of \citet{tiscareno03} (a
reproduction of their Fig.\ 10a). {\bf Bottom:} Centaur detection
probability based on the \citet{tiscareno03} Monte Carlo simulation
(reproduction of their Fig.\ 10b). The black line shows an exponential
fit to their data. Combining the two curves suggests that $< 4$\% of
Centaurs have been detected.}
\label{fig:detectability}
\end{figure}

The first estimate of the total number of Centaurs comes from the
simulations of \citet{tiscareno03}, who investigated the dynamical
evolution of the observed Centaurs over 100~Myr. The top panel of Figure
\ref{fig:detectability} shows their computed time-averaged eccentricity
distribution. \citet{tiscareno03} also estimated the detection fraction
of Centaurs as a function of eccentricity, which is reproduced in the
bottom panel of Figure \ref{fig:detectability}. The detection fraction
estimate holds for Centaurs with $R \geq 30$~km.  Multiplying the
eccentricity distribution with a fit to the detectability function
(black line in the bottom panel of Figure \ref{fig:detectability}) and
summing over the 0-1 eccentricity range yields an estimate of $f_{\rm
det} = 3.7$\% for the fraction of Centaurs with $R \geq 30$~km that have
been detected. The total number of large Centaurs with $R \geq 30$~km is
then is $\sim N_{\rm obs} / f_{\rm det}$, where $N_{\rm obs} = 53$ is
the number of Centaurs that had been discovered when the
\citet{tiscareno03} calculations were performed.

The next step in determining the total number of Centaurs is to find the
radius of the largest Centaur. For $R \geq 30$~km,
\begin{equation}
N_{\geq}(R) = \frac{N_0}{\zeta - 1} \left ( \frac{R}{R_0} \right
)^{1-\zeta} .
\label{eq:cumnum}
\end{equation} 
In Equation \ref{eq:cumnum}, $N_{\geq}(R_{\rm max}) = 1$, $N_{\geq}(30
\; {\rm km}) = N_{\rm obs} / f_{\rm det} = 1432$, and $\zeta = 4$
\citep[e.g.][]{trujillo01, fraser08, minton12, schlichting13}, so that
$R_{\rm max} = 338$~km.  An estimate of the total number of Centaurs
then follows, given an analytical form for the differential size
distribution $dN/dR$. \citet{schlichting13} find a KBO size
distribution of the form $dN/dR \propto R^{-\zeta}$, where $\zeta = 2$
for 10~km$\leq R \le 30$~km; $\zeta = 5.8$ for 2~km$\leq R \le 10$~km;
and $\zeta = 2.5$ for 0.1~km$\leq R \le 2$~km.  We set a lower limit of
$R = 1$~km to the size of Centaurs considered here, which is justified
because the mass contained in the smallest bodies is negligible unless
$\zeta \geq 4$.  The \citet{schlichting13} conclusion that $\zeta < 4$
for the smallest bodies is supported by sky brightness measurements,
which rule out $\zeta \geq 3.4$ for $R < 1$~km \citep{kenyon01,
ichikawa11}.  The size distribution computed based on the
\citet{tiscareno03} maximum Centaur-size estimate contains $2.8 \times
10^7$ comets with $R \geq 1$~km, and is shown in the top panel of Figure
\ref{fig:sizedist} (red curve). The distribution agrees well with the
results of \citet{sheppard00}, who predict about 100 Centaurs with radii
above 50~km. However, the number of small bodies is an order of
magnitude lower than the \citet{disisto07} estimate of $\sim 2.8 \times
10^8$ Centaurs with radii above 1~km.



Other estimates of the total number of Centaurs come from radius
measurements of Centaurs and KBOs. The most conservative estimates come
from assuming that Chariklo, the largest observed Centaur, is in fact
the largest Centaur in the Solar System. (It is highly likely that the
largest Centaur has not been observed, given that the detection
probability is extremely low for even moderately eccentric orbits.)
Chariklo radius estimates range between 118~km and 151~km
\citep{fornasier13, stansberry08, groussin04, altenhoff01, jewitt98}.
The green and black curves in Figure \ref{fig:sizedist} show size
distributions where the largest body takes on the maximum and minimum
observational estimates of Chariklo's radius, respectively. Finally,
Figure \ref{fig:sizedist} shows a size distribution that is optimistic
about the size of the largest body, with $R_{\rm max} = 458.5$~km, the
maximum measured radius of the Plutino Orcus (blue curve). Larger TNOs
such as Quaoar and Pluto have higher densities that suggest
differentiation, whereas Orcus' bulk density is more consistent with the
undifferentiated comet population. It is plausible, then, that Orcus
represents a transition object between Kuiper Belt comets/Centaurs and
true dwarf planets. Note that this ``optimistic'' size distribution
predicts only a factor of 2.5 more Centaurs than using the
\citet{tiscareno03} results, but brings the number of small
bodies closer to the predictions of \citet{disisto07}.

The bottom panel of Figure \ref{fig:sizedist} shows the cumulative mass
function $M_<(R)$, the mass of Centaurs with radii less than a given
value. For each size distribution from the top of Figure
\ref{fig:sizedist}, three possible Centaur densities are considered.
The highest density is the maximum inferred value for Orcus
(1.53~g~cm$^{-3}$, \citet{stansberry12}, dashed lines). The solid lines
show mass functions with the bulk density of KBO Salacia, which has the
lowest bulk density measurement of any TNO (1.16~g~cm$^{-3}$,
\citet{stansberry12}). Finally, the smallest Centaurs may have very low
densities comparable to comet 67P/Churyumov-Gerasimenko, $<
1$~g~cm$^{-3}$ \citep{sierks15}. The dash-dot lines show cumulative mass
distributions with $\rho = 0.5$~g~cm$^{-3}$. All Centaurs are given the
same density regardless of size, though in reality the larger bodies
probably have higher densities.

It is important to note that the abundance of large Centaurs may be out
of equilibrium with the source KBO population, due both to stochastic
effects and to the possibility that the diffusion rate from the Kuiper
Belt to the inner Solar System is a function of size. Since impacts by
100-km objects onto ice giants occur less than once per 10~Myr, these
objects may be safely neglected from any discussion of the steady-state
ice-giant luminosities. However, the size distributions presented here
are anchored by the large objects, so disequilibrium from the source KBO
population could lead to large uncertainties the number of small bodies.




\section{Impact-generated luminosity}
\label{sec:luminosity}

Given the Centaur size distribution, impact speed, and radiative time
constants of the ice giant atmospheres, we can compute the
impact-generated luminosity of ice giants as a function of time.  The
impact speed is $v_{\rm imp}^2 = v_{\rm esc}^2 + (e v_{\rm orb})^2$,
where $v_{\rm esc}$ is the planet's escape speed, $e$ is the Centaur's
eccentricity, and $v_{\rm orb}$ is the Centaur's orbital speed.  For
Centaurs crossing the orbits of Uranus and Neptune, $v_{\rm orb} \ll
v_{\rm esc}$, and impacts occur roughly at escape speed. Each impact
converts Centaur kinetic energy into heat, which the ice giant
atmospheres radiate away over a characteristic timescale.

Radiative time constants $\tau_{\rm rad}$ used here are 95 years for
Uranus and 105 years for Neptune.  \citet{conrath90} computed radiative
time constants based on profiles of $Q_{\rm IR}$, the infrared energy
deposited per unit volume and time at each level of the planet
atmosphere. $Q_{\rm IR}$ is usually negative, indicating cooling, and is
dominated by methane emission. To compute $\tau_{\rm rad}$,
\citet{conrath90} first constructed model atmospheres by imposing the
latitudnally dependent temperature measured by Voyager data as a
boundary condition, then perturbed the temperature profile at each
atmosphere depth to compute the change in $Q_{\rm IR}$. Changes in the
composition of ice giant atmospheres, such as aerosol darkening
predicted to correspond with the 11-year solar cycle \citep{baines90,
hammel07}, could alter the radiative time constants; how much is not
clear as the \citet{conrath90} calculation has not been updated based on
new data. Since radiative time constants become independent of
atmospheric pressure for $P > 0.5$~bar (the boundary between the
radiative and convective layers according to \citet{conrath90} and
\citet{depater10}) and even the smallest impactors considered ($R =
1$~km) would penetrate deeper than the 1-bar level, impact-generated
luminosities are assumed not to be sensitive to the penetration depth of
the impactors.

We use a Monte Carlo simulation of successive ice giant impacts from
Centaurs with radii chosen randomly from the size distributions plotted
in Figure \ref{fig:sizedist}. The Monte Carlo simulation reveals how
often Centaur impacts can bring Neptune's internal luminosity (excluding
reprocessed energy from the Sun) up to the observed value of $3.64
\times 10^{22}$~erg~s$^{-1}$. A sequence of impacts meets the criteria
for a Poisson process: gravitational interactions between Centaurs are
negligible, so the probability of an impact is independent of previous
impact history, and the likelihood of an impact in a given amount of
time increases with the length of the time interval under consideration.
Impacts are therefore modeled as a Poisson process with rate parameter
$\lambda = 1/\tau_{\rm imp}$, where $\tau_{\rm imp}$ is the mean time
interval between impacts, and the distribution of time intervals between
impacts is exponential. Only impactors larger than 1~km are included in
the simulation. Based on their models of diffusion from the Kuiper Belt,
\citet{levison97} calculated that there should be $1.2 \times 10^7$
ecliptic comets (Centaurs, Halley-type comets and Jupiter-family comets
combined\footnote{The size distributions based on measurements of
Chariklo and Orcus do not encode any dynamical information about the
impactors, so that they may represent Jupiter-family comets or
Halley-type comets as well as Centaurs. No correction needs to be made
to the total number of objects to be consistent with the
\citet{levison97} impact rates. Of the estimates of the total number of
Uranus and Neptune impactors in Section \ref{sec:sizedist}, only the one
based on the \citet{tiscareno03} detectability simulation is restricted
to true Centaurs with $Q > 30.2$~AU. However, since \citet{levison97}
find that true Centaurs are far more common than shorter-period comets,
we also make no correction to the total number of impactors from this
size distribution.}), assuming a maximum radius of $R_{\rm max} =
200$~km and a minimum radius $R_{\rm min}$ in the range 0.5-2~km.
\citet{levison97} find impact rates of $\lambda_0 = 1.2 \times
10^{-3}$~yr$^{-1}$ for Neptune and $\lambda_0 = 1.3 \times
10^{-3}$~yr$^{-1}$ for Uranus. Rate parameters are scaled to the
\citet{levison97} results according to the number of Centaurs with radii
over 1~km, such that
\begin{equation}
\lambda = \lambda_0 \left ( \frac{N_{\geq}({\rm 1 \; km})}{1.2 \times 10^7} \right) . 
\label{eq:lambda}
\end{equation}

The simulation is built on the simplifying assumption that impact energy
is deposited instantaneously and then redistributed throughout the
atmosphere immediately. Subsequent to an impact, Newtonian cooling
governs impact-generated planet luminosity as a function of time.
Luminosity is then determined by both the impact energy $E_{\rm imp}$
and the atmospheric radiative time constant:
\begin{equation}
L = (L_{\rm imp}+L_0) e^{-(t - t_{\rm imp})/\tau_{\rm rad}}
\label{eq:luminosity}
\end{equation}
\begin{equation}
L_{\rm imp} = E_{\rm imp} / \tau_{\rm rad}
\label{eq:limp}
\end{equation}
In Equation \ref{eq:luminosity}, $t_{\rm imp}$ is the time of the most
recent impact and $L_0$ represents any residual luminosity left over
from previous impacts. We assume 100\% conversion of impactor kinetic
energy to heat, though in reality some energy will be lost to
sublimation and ablation of the impactor. As the goal of this experiment is
to assess how much of the ice giant energy balance can be explained by
impacts, no other source of internal energy (such as gravitational
contraction or radioactive decay) is included.

\begin{figure}
\centering
\begin{tabular}{cc}
\includegraphics[width=0.5\textwidth]{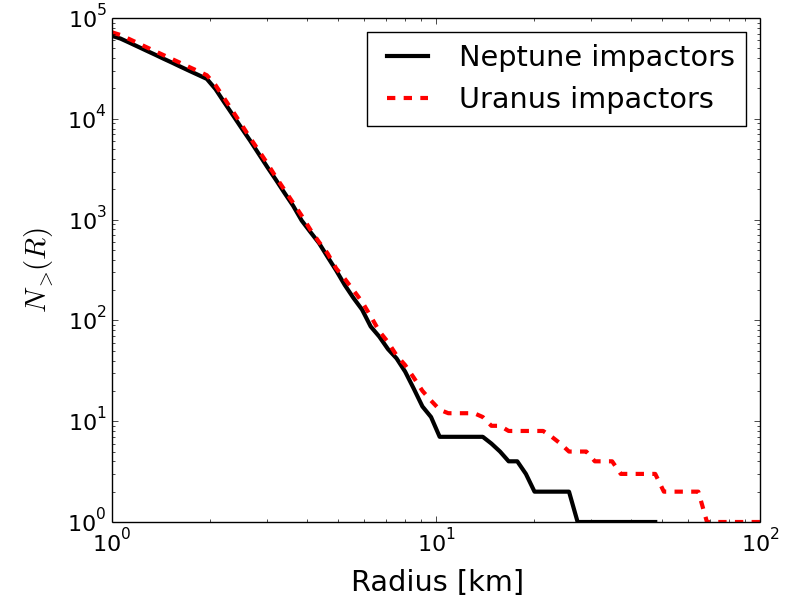}
\includegraphics[width=0.5\textwidth]{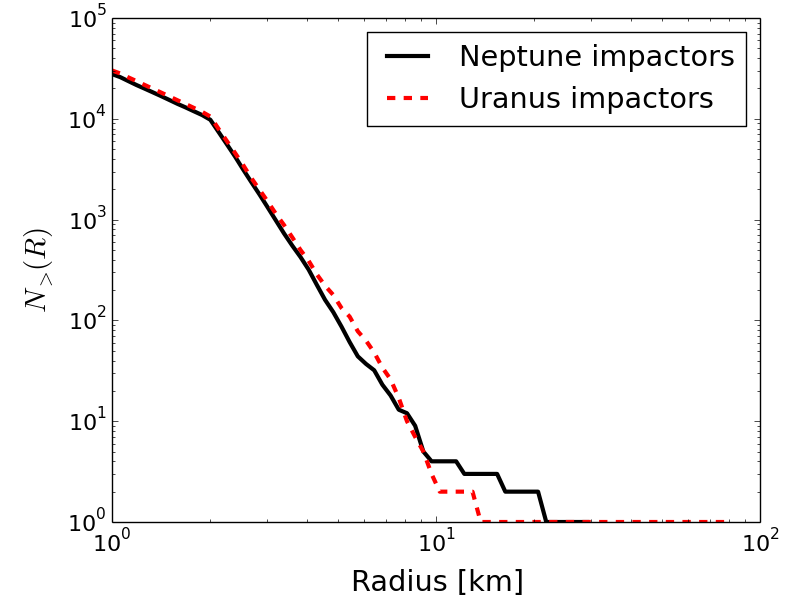}
\end{tabular}
\caption{Sample size distributions of impactors chosen from 1~Myr Monte
Carlo simulation. 
Stochastic behavior appears for larger Centaur radii.
}
\label{fig:mcsizedist}
\end{figure}

The simulation proceeds as follows: (1) a random set of impact time
intervals is drawn from an exponential distribution.
(2) For
each impact, a Centaur radius is randomly selected from the size
distribution under consideration. Figure \ref{fig:mcsizedist} shows two
realizations of randomly chosen impactor size distributions.
(3) The time evolution of the planet's luminosity is calculated.  At
each impact time, the luminosity $L_{\rm imp}$ is added to any remaining
luminosity from previous impacts $L_0$ and a Newtonian cooling process
starts (Equations \ref{eq:luminosity} and \ref{eq:limp}). (4) The
cumulative probability distribution of planet luminosity is calculated.

\section{Results and Conclusions}
\label{sec:results}

\begin{figure}
\centering
\includegraphics[width=0.8\textwidth]{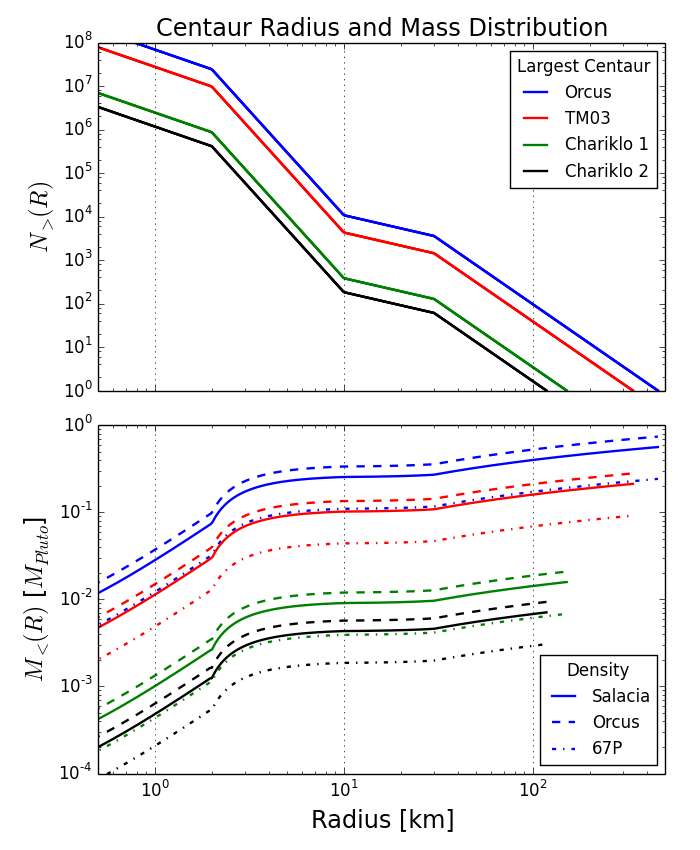}
\caption{{\bf Top:} Size distribution $N_>(R)$, the number of Centaurs
with radii greater than size $R$. Colors show size distributions
corresponding to different assumed sizes of the largest Centaur.  For
the red curve marked TM03, the largest Centaur radius is calculated
self-consistently using the Centaur detection probability of
\citet{tiscareno03}, which suggests that 3.7\% of the Centaurs with $R >
30$~km have been detected. The blue curve marked ``Orcus'' shows a size
distribution for which the largest Centaur has the diameter of Orcus,
917~km. The green and black lines labeled Chariklo~1 and Chariklo~2 show
size distributions with the largest Centaur diameter given by the
minimum and maximum measured values for Chariklo. {\bf Bottom:}
Mass distribution $M_<(R)$, the total mass of Centaurs with
radii less than size $R$. The color scheme is as above. Line
styles denote assumptions about Centaur density. Orcus has a
maximum measured density of 1.5~g~cm$^{-3}$, Salacia has a
maximum measured density of 1.3~g~cm$^{-3}$, and comet 67P
(Churyumov-Gerasimenko) has a density of 0.5~g~cm$^{-3}$.}
\label{fig:sizedist}
\end{figure}

\begin{figure}
\centering
\begin{tabular}{cc}
\includegraphics[width=0.49\textwidth]{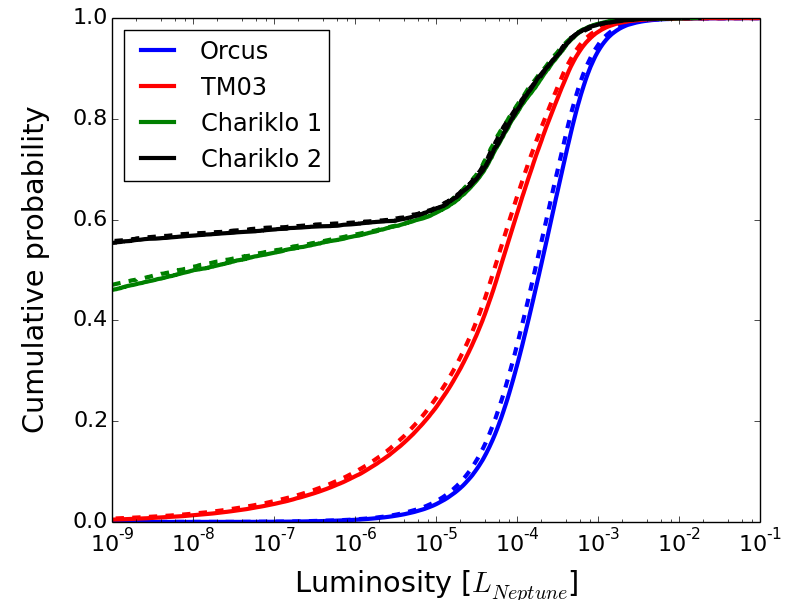}
\includegraphics[width=0.49\textwidth]{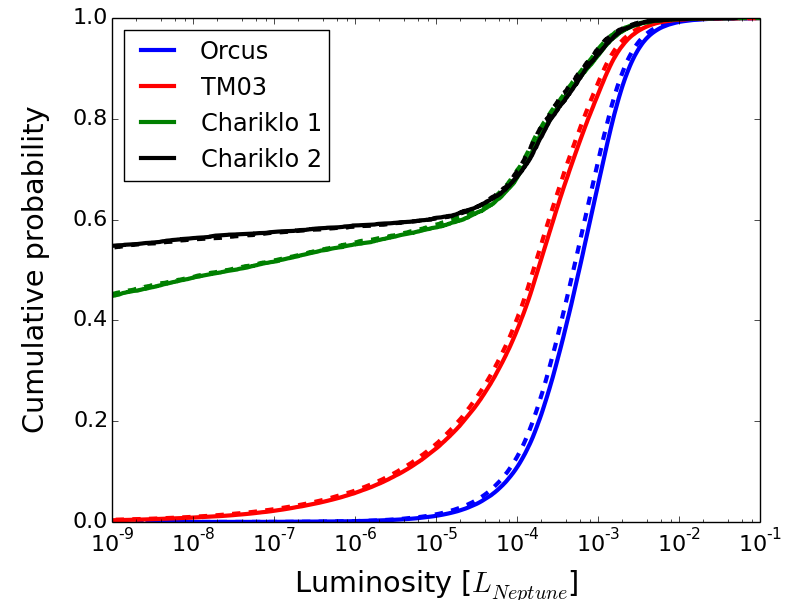}
\end{tabular}
\caption{Cumulative probability distribution of impact-generated planet
luminosity for each of the four size distributions under consideration.
In the left panel, the model Centaurs have the same density as Comet
67P, $\rho = 0.5$~g~cm$^{-3}$. In the right panel, the assumed Centaur
density is $\rho = 1.5$~g~cm$^{-3}$.  The color scheme is the same as in
Figure \ref{fig:sizedist}. Solid lines show Neptune luminosities and
dashed lines show Uranus luminosities. The minimum luminosity produced
by an impact is $L_{\rm imp} = 2 \times 10^{18}$~erg~s$^{-1}$. Lower
values indicates long time intervals between impacts.}
\label{fig:luminosityprob}
\end{figure}

Figure \ref{fig:luminosityprob} shows probability distributions of
planet luminosity from a 10-Myr simulation, for all four size
distributions shown in Figure \ref{fig:sizedist}, and for Centaur
densities of 0.5~g~cm$^{-3}$ (left) and 1.5~g~cm$^{-3}$ (right). For
only the two most populous size distributions did the simulations record
impacts that produce luminosities of order Neptune's observed value,
$3.64 \times 10^{22}$~erg~s$^{-1}$---and then extremely infrequently. If
the total Centaur mass is of order Pluto's mass (see blue dashed curve
in the bottom panel of Figure \ref{fig:sizedist}), an impact that yields
a luminosity of $L_{\rm imp} > 10^{22}$~erg~s$^{-1}$ occurs every
0.6~Myr. The largest luminosity generated by a single impact that the
simulations produced was $L_{\rm imp} = 3.3 \times
10^{23}$~erg~s$^{-1}$, from a 13-km impactor with $\rho =
1.5$~g~cm$^{-3}$.

Based on current understanding of Centaur dynamics and radii, it is
clear that impacts cannot explain the ice giants' energy balance. Not
only are energetic impacts far too rare to provide Neptune's luminosity,
but \citet{levison97} show that Uranus and Neptune experience impacts at
almost the same rate. Furthermore, their atmospheres radiate away energy
on similar timescales. Only after a rare giant impact onto
one planet are their impact-induced luminosities are substantially
different.  Instead, Uranus and Neptune's different energy balances are
probably the result of a structural dichotomy between the two planets
\citep{nettelmann13}, though shape and rotation data are not yet precise
enough to elucidate exactly how the ice giant interior structures
differ.

There are two circumstances in which it may be worth revisiting the
connection between impacts and ice giant energy balance. First, a
re-calculation is warranted if observers discover a significant
population of Neptune-crossing comets that have not been well
represented in dynamical simulations (i.e. high-inclination objects).
With a factor of 100 increase in Centaur abundance over and above the
most populous Centaur size distribution considered here, impacts would
contribute substantially to Neptune's steady-state energy balance.
Furthermore, high-inclination Centaurs from the Oort cloud have longer
dynamical lifetimes than those with low inclinations
\citep[e.g.][]{disisto07, brasser12, volk13}, which may indicate that
this work underestimates the steady-state Centaur abundance. The number
of low-inclination Centaurs sourced from the scattered disk and
cold Kuiper Belt is not likely to increase much as both populations have
been well characterized \citep[e.g.][]{adams14}. The highest estimate of
the number of Centaurs \citep{disisto07} is a mere factor of four higher
than the maximum number considered here, and \citet{horner04} predict
only $4.4 \times 10^4$ Centaurs larger than 1~km. We do not
consider a substantial upward revision to the total number of
Centaurs likely.
Second, if models of the KBO size distribution are revised to include a
significantly higher proportion of $\sim 10$-km objects, median impact
energies would produce luminosities of order Neptune's current value.
However, for impact rates calculated by \citet{levison97}, changing the
KBO size spectrum to favor 10-km objects could push Uranus' luminosity
well over observed values. In any case, planetesimal-growth simulations,
occultation statistics, and crater size measurements are converging in a
way that make substantial revisions to the 10-km KBO abundance unlikely
\citep[e.g.][]{schlichting09, kenyon12, minton12, schlichting13}.

One possibly fruitful avenue for future research is investigating the
connection between impacts and seismic disturbances or storms. A comet
of radius 1~km would deposit $1.2 \times 10^{28}$~erg of energy locally
over a timescale of $\sim 1000$~seconds, based on the Shoemaker-Levy
Jupiter impact models of \citet{harrington01}. \citet{marley94}
demonstrated that impacts of energies $10^{28}$~ergs and higher would
produce detectable temperature fluctuations exceeding 1~K on Jupiter due
to the excitation of seismic waves.
However, impacts by 1~km objects are still too infrequent---once per
hundred years, even for generous estimates of the number of Centaurs
such as those of \citet{disisto07}---to make ice-giant seismology a
reasonable observing campaign. Detection of seismic activity would be
serendipitous. It is faintly possible, however, that sedimentation of
particles ablated from an impactor could be observable. In Keck NIRC2
adaptive optics observations of Uranus on 12 August 2004,
\citet{sromovsky05} noted three low surface-brightness features that
appeared and disappeared on timescales of 0.3-1.5~hours. Based on
observations by \citet{carlson88}, \citet{sromovsky05} suggest that
sedimentation of 300-$\mu$m particles caused by vertical motion of a
cloud could produce the rapid disappearance of the faint features. While
vertical shear acting on clouds is the most likely reason for small
particles to descend in Uranus' atmosphere, we suggest cautiously
considering impactors a source of settling particles.  Extending the
\citet{schlichting13} size distribution down to its lowest radius limit
of 0.01~km yields a total of $> 10^{11}$ Centaurs, even assuming the
smallest maximum Centaur size considered here (118~km; minimum measured
Chariklo radius). At minimum, tens of 0.01~km-size objects would impact
each ice giant per year. Given that Uranus has recently shown
record-breaking storms \citep{depater15}, despite its equinox passage
possibly decreasing convective activity, it is worth considering whether
impacts may trigger local storms.

Funding for this project was provided by NSF grant AST-1055910. The idea
that impacts may be responsible for Neptune's high luminosity grew out
of a discussion in the upper-division Solar System Astronomy course at
University of Texas in Fall 2011. I thank Andy Liao for asking questions
that led me to perform this calculation. I am grateful to Renu Malhotra
and Matthew Tiscareno for sharing data from their 2003 paper, and for
comments that improved this work. Thoughtful reviews from two anonymous
referees improved the manuscript. Advanced undergraduate and graduate
students at both Univeristy of Texas and University of Delaware have
performed an order-of-magnitude version of this calculation on their
homework assignments. Please contact me if you would like to use the
assignment in your course.





\end{document}